# A Silicon-Singlet Fission Parallel Tandem Solar Cell Exceeding 100 % External Quantum Efficiency

Luis M. Pazos[2,1], Ju Min Lee,[1] Anton Kirch,[2] Maxim Tabachnyk,[2] Richard H. Friend,[2] and Bruno Ehrler[1]

(1)     FOM Institute AMOLF, Center for Nanophotonics, Science Park 104, 1098 XG Amsterdam, The Netherlands

(2)     Cavendish Laboratory, University of Cambridge, JJ Thomson Avenue, Cambridge CB3 0HE, UK

## Abstract

Silicon solar cells dominate the solar cell market with record lab efficiencies reaching almost 26%. However, after 60 years of research, this efficiency saturated close to the theoretical limit for silicon, and radically new approaches are needed to further improve the efficiency. Here we present parallel-connected tandem solar cells based on down-conversion via singlet fission. This design allows raising the theoretical power conversion efficiency limit to 45% with far superior stability under changing sunlight conditions in comparison to traditional series tandems. We experimentally demonstrate a silicon/pentacene parallel tandem solar cell that exceeds 100% external quantum efficiency at the main absorption peak of pentacene, showing efficient photocurrent addition and proving this design as a realistic prospect for real-world applications.

Conventional single-junction solar cells are limited in efficiency to about 34%, mainly due to non-absorbed below-bandgap photons and the loss of energy via thermalization of high-energy electron-hole pairs. This limit is called the Shockley-Queisser limit.[1] Singlet fission is a down-conversion process in organic semiconductors that spontaneously converts one high-energy spin-singlet electron-hole pair (exciton) into two spin-triplet excitons.[2] Each triplet exciton carries half the energy of the initial singlet exciton. Utilized in solar cells, this process could lift the theoretical limit of a single junction[3,4] when combined with a lower-bandgap semiconductor.

In previous work, we and others have shown successful examples which incorporated pentacene as the singlet fission sensitizer for lead chalcogenide quantum dots[5–7] or amorphous silicon.[8] Here we use a novel architecture, combining a conventional monocrystalline silicon solar cell with a pentacene cell connected electrically in parallel. In such a parallel-tandem architecture the efficiency of silicon photovoltaics can be enhanced with singlet fission by potentially doubling the current obtained from high-energy photons. Tandem solar cells already overcome[9] the single-junction Shockley-Queisser limit by stacking two or more solar cells with a different bandgap in series such that light passes the high-bandgap material before it reaches the lower-bandgap sub-cell(s) (see Figure 1 (A)). In this configuration, steady-state is reached when the voltages of the sub-cells add, and the currents match. A mismatch between the current generated by each sub-cell forces a shift on their corresponding operation voltages from their optimal points. For this reason a mismatch in current leads to a drop in efficiency. The design and manufacturing for obtaining current-matching is challenging and very costly, and this match cannot be maintained as the solar spectrum changes, particularly under diffuse illumination. As a result, tandem cells are only viable in locations with direct illumination, and are restricted to niche markets, for example concentrator solar cells.

In contrast, when two solar cells are electrically connected in parallel, they operate at the same voltage and the currents add. Voltage scales only logarithmically with light intensity rather than linearly,[10] hence, as we show here, voltage matching is far easier to achieve for changing sunlight conditions as compared to current-matching, and more robust against fabrication constraints and materials mismatch. For conventional solar cells the voltage is mostly determined by the bandgap, hence a two-bandgap parallel tandem configuration could not achieve voltage-matching without complex contacting schemes combining different numbers of sub-cells. However, when the high-bandgap sub-cell is a singlet fission solar cell, voltage matching is possible in a single, two-terminal solar cell.

The singlet fission down-conversion from one high-energy exciton to two lower-energy excitons allows the current from the solar cell to double while the voltage is reduced by about a factor of two compared to a conventional cell of the same bandgap. On its own, such a singlet fission cell allows no higher conversion efficiency than a conventional cell, since the extracted power (current × voltage) remains constant for this single-bandgap cell. However, connecting a high-bandgap singlet fission solar cell in parallel with a conventional solar cell of lower bandgap can form a voltage-matched two-bandgap system (see Figure 1 (B)). When the two sub-cells are optically connected in series, the light can pass both layers successively resulting in a solar cell with theoretical efficiencies greater than those for any single-junction solar cell. To find the limiting efficiency we use a detailed balance model following Shockley and Queisser[1] with modifications for singlet fission similar to Hanna and Nozik.[4] The main difference for parallel tandem solar cells compared to conventional series-tandem cells is that the generation and recombination current of both sub-cells adds for the complete tandem cell (see Supplementary Information S2 for details). In a series tandem the current of both sub-cells equilibrates and the voltages are added. Figure 1 (C) and (D) show the limiting efficiency for series and parallel tandem solar cells respectively. The highest theoretical efficiency that can be reached is around 45% in both cases for a low-bandgap semiconductor of 0.94 eV for the bottom cell and 1.60 eV for the top cell in the case of a series tandem cell, while for the parallel tandem cell the optimum combination is 0.95 eV for the low bandgap cell and 1.64 eV for the singlet fission cell (see also Table 1).

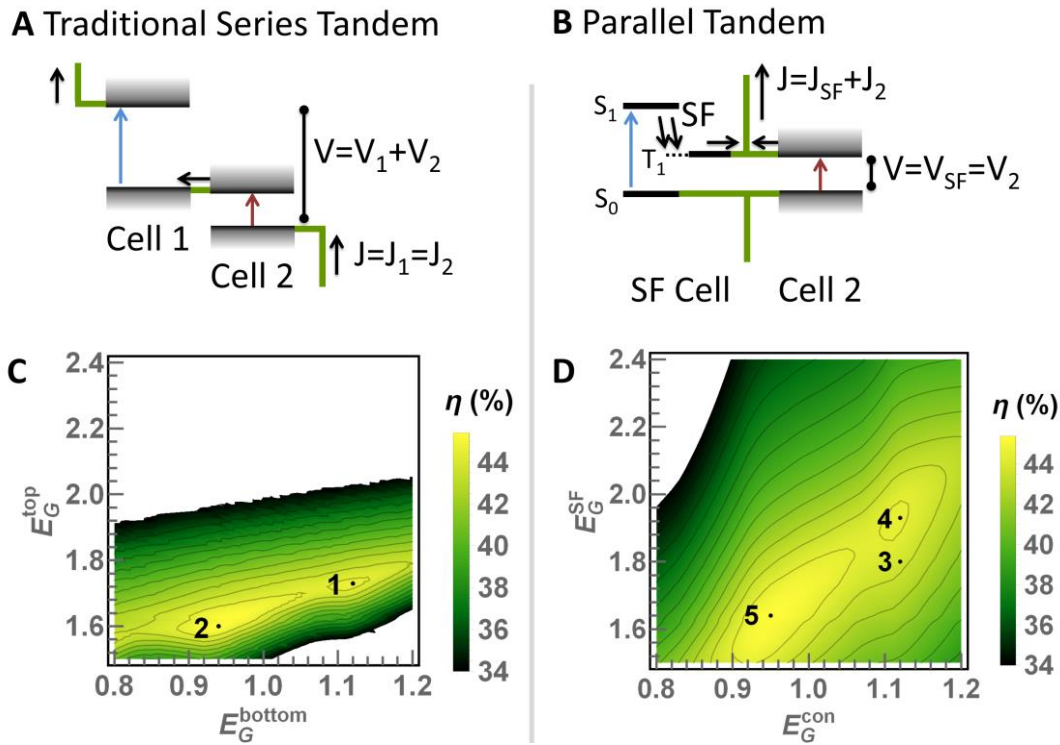

Figure 1. (A) Traditional tandem solar cell, electrically and optically connected in series. (B) Parallel singlet fission tandem cell. The singlet fission down-conversion facilitates voltage matching by producing two low-energy excitations from one high-energy photon. (C) Detailed-balance calculation for the efficiency limit of series tandem solar cells with varying bandgaps for both the top and bottom cell. The numbers represent the cases displayed in Table 1. (D) The efficiency limit for the parallel tandem solar cell where the conventional semiconductor has bandgap $E_G^{bottom}$ and the singlet fission semiconductor $E_G^{SF}$. Only values exceeding the single-junction Shockley-Queisser limit are shown.

Contrary to the series tandem cell, the parallel tandem configuration does not require current matching of the two sub-cells. The voltage needs to be matched as closely as possible, but since changes in bandgap lead to smaller changes of voltage than current (see Supplementary Information S3), high efficiencies in a parallel tandem solar cell are easier to achieve for a broader range of materials with different bandgaps, without compromising the efficiency by incomplete absorption. This difference is illustrated in Figure 1 (D) where we plot the combination of bandgaps exceeding the single-junction limit.

Table 1. Maximum theoretical efficiency of various single-junction and parallel tandem solar cells calculated using detailed balance.

| Configuration | Figure 1 | Bandgap(s) | Limit $\eta$ |
|---|---|---|---|
| **Ideal single junction** | | 1.34 eV | 33.8% |
| **Single junction silicon** | | 1.12 eV | 33.2% |
| **Series ideal top and silicon bottom cell** | (c) 1 | 1.73 eV / 1.12 eV | 44.7% |
| **Series ideal top and bottom cell** | (c) 2 | 1.60 eV / 0.94 eV | 45.4% |
| **Parallel pentacene/silicon** | (d) 3 | 1.80 eV / 1.12 eV | 43.4% |
| **Parallel ideal SF material/silicon** | (d) 4 | 1.95 eV / 1.12 eV | 44.3% |
| **Parallel ideal SF material and bottom cell** | (d) 5 | 1.64 eV / 0.96 eV | 45.2% |

Crucially, the efficiency limit of the parallel tandem cell is less affected by changing spectral conditions. The spectral shape can change due to the angle between the cell and the sun, atmospheric conditions, time of the day, cloud coverage etc.; such changes alter the relation between direct and diffuse sunlight. To illustrate the stability of the tandem solar cell efficiency against changes in the spectrum we calculate the limiting efficiency for a series- and a parallel tandem solar cell as well as a single-junction cell, all three optimized for standard AM1.5G illumination, when the spectral shape changes. Figure 2 (top) shows the average limiting efficiency for spectra measured near Rotterdam (Netherlands) for each day of 2014. Even though the series tandem cell has a much higher efficiency limit under AM1.5G conditions when compared to the single-junction, its average efficiency limit under the real spectra in a cloudy country clusters around the same efficiency limit as the single-junction cell. This arises because the ratio of diffuse and direct light changes during the day, and these two components of sunlight have a very different spectral shape (see inset Figure 2). As a result one of the sub-cells in the tandem stack receives less light than the other, resulting in a mismatch in current. The series tandem solar cell is always limited by the lowest of the currents, making its efficiency limit vulnerable to the change in the ratio of diffuse vs. direct light (see Supporting Information S4). In contrast, parallel tandem solar cells are far more stable against those changes, and the efficiency advantage over the single-junction solar cells remains almost independent of the incoming spectrum. The power output of an ideal series tandem, parallel tandem, and single-junction solar cell under the same spectra is shown in the bottom part of Figure 2. The difference between the series tandem and parallel tandem cells is smaller, because under clear-sky conditions, when the efficiency difference is marginal, the irradiated power is largest. However, there is still a very significant difference in the possible power output, with a parallel tandem cell potentially providing 33% more power over the year than a single-junction cell, compared to only 18% more power in case of the series tandem cell.

We note that the analysis presented here holds also for parallel connection of any voltage matched architecture, being that achieved *via* down-conversion, up-conversion or combinations of series-connected sub-cells arranged to achieve voltage matching from cells with different $V_{OC}$[11]; this may be an effective approach for two-terminal lead halide perovskite cells in parallel tandem with silicon. We also note that in realistic architectures, the fill factor is sub-ideal, which makes the efficiency slightly less sensitive to changes in illumination spectrum.

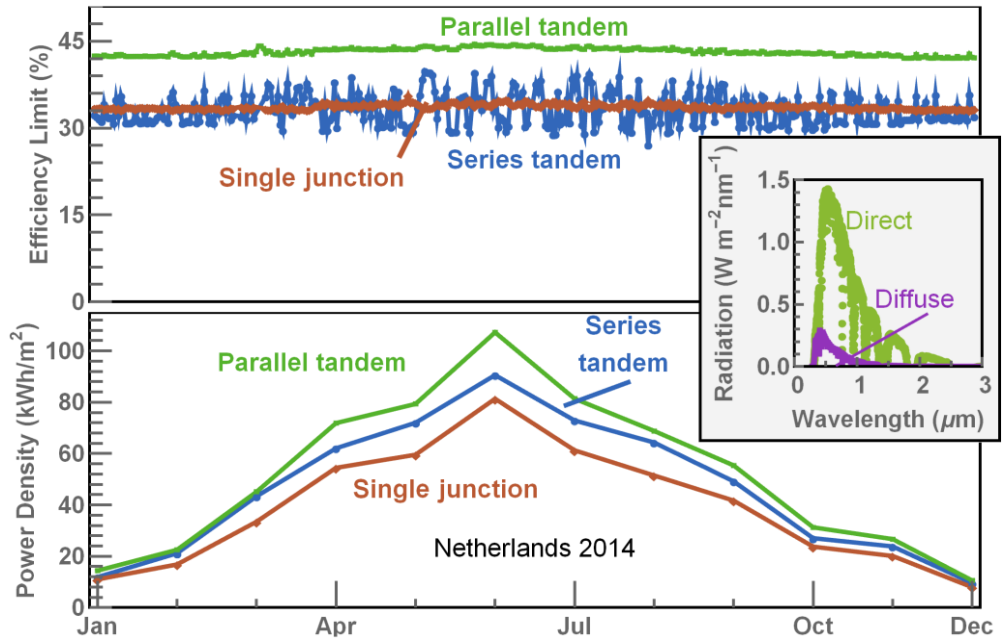

**Figure 2. Influence of spectral variation on the average daily efficiency limit (top), with spectra experimentally measured in the Netherlands at different days in 2014, and on the total power density for an ideal cell per month of 2014 (bottom). The inset shows the direct and diffuse part of the AM1.5G standard solar spectrum.**

Our implementation of the parallel tandem solar cell consists of a mono-crystalline silicon solar cell as the low bandgap sub-cell and a pentacene/ $C_{60}$ fullerene heterojunction solar cell as the high-bandgap singlet fission sensitized sub-cell (see Figure 3 (A)). Every high-energy singlet exciton (1.8 eV) in pentacene is rapidly converted[12] into two triplet excitons of lower energy (0.9 eV[7]), leading to a triplet exciton yield of 200%.[13] Those triplet excitons can be efficiently dissociated at the interface between pentacene and $C_{60}$. As a result, pentacene/$C_{60}$ solar cells have shown very high external and internal quantum efficiencies, exceeding 100% and approaching 200% respectively.[13–16] Surprisingly high open-circuit voltages have been observed for pentacene/$C_{60}$[13] and pentacene/quantum dot[7] solar cells, suggesting a reasonably low triplet exciton binding energy in pentacene. Hence, here we use pentacene as the singlet fission sensitizer to demonstrate the potential of utilizing the parallel tandem architecture to combine a singlet fission material with silicon. Table 1 shows that the limiting efficiency for the pentacene/silicon tandem (43.9%) is relatively close to the maximum efficiency and about a third above the limit for a single-junction solar cell. Other singlet fission sensitizers which have an even better suited triplet exciton energy to match the silicon bandgap and voltage, such as terrylenes,[17] and perylenediimides,[18] have not been applied in solar cells yet, and singlet fission solar cells made from tetracene[19] and its derivatives,[20] have lower theoretical efficiency in a parallel tandem cell.

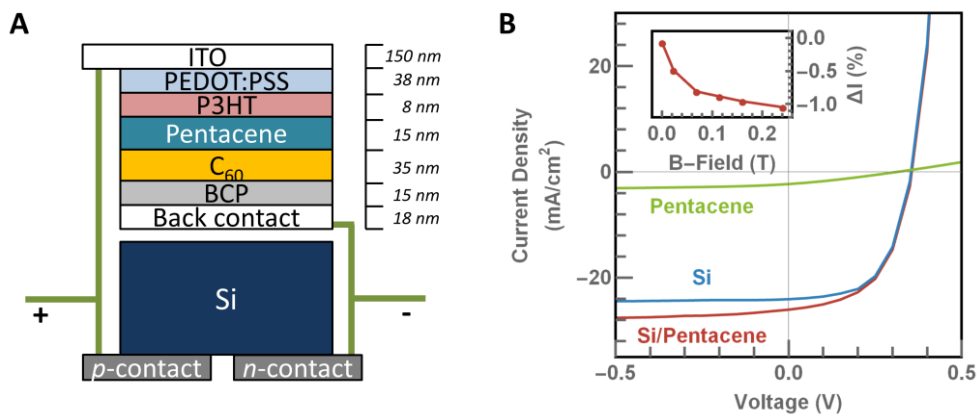

**Figure 3.** (A) Device architecture. The light, incident from above, is split into the high-energy part absorbed by the pentacene sub-cell, and the low-energy part absorbed by the silicon cell. (B) In the parallel tandem architecture, the currents from both sub-cells add up at every voltage. The inset shows the change in current from the pentacene cell under a magnetic field.

We build the singlet fission solar cell following work by Congreve et al.[13] The pentacene device is made of an ITO electrode, 38 nm PEDOT:PSS, 8 nm of P3HT, 15 nm of Pentacene, an electron accepting layer of 35 nm of $C_{60}$, 15 nm of bathocuproine (BCP). We use a transparent ITO front-contact and a semi-transparent back-contact (LiF 1 nm/Al 1.5 nm/Ag 15 nm).[21] Thus we can place the singlet fission cell directly in front of the silicon solar cell in such a way that high-energy photons ($E_{hv}$ > 1.8 eV) are absorbed in the singlet fission cell, while low energy photons (1.1 eV < $E_{hv}$ < 1.8 eV)) reach the silicon cell underneath.

The combination of both solar cells produces the sum of the current of the two sub-cells. The current-voltage (IV) characteristics of the two sub-cells measured individually (already in the device stack) compared to the case where both cells are connected in parallel demonstrates this current addition (Figure 3 (B)). The magnetic field dependence of the photocurrent (see inset Figure 3 (B)) from the pentacene cell shows the typical decrease in photocurrent under high magnetic fields. This current corresponds to the change in singlet fission efficiency in pentacene[22] and confirms that the photocurrent from the pentacene cell predominantly originates from fission-generated triplet excitons.

The current addition is seen more clearly in the photon-to-electron conversion efficiency (external quantum efficiency, EQE, Figure 4 (A)) where the pentacene cell contributes to the current generated by the silicon cell, reaching a peak of 65% EQE at 1.85 eV (red trace). The contacts of the pentacene cell absorb around 30% of the light, and parasitic losses and reflection at the air/glass interface further reduce the amount of light reaching the silicon cell. Where the pentacene absorbs, even less light reaches the silicon cell (blue trace). Nevertheless, the pentacene peaks (green trace) are clearly visible in the EQE of the combined cell (red trace), demonstrating the contribution of carrier multiplication to the photocurrent.

To illustrate the potential of our technology we measure the tandem cells in a slightly modified configuration where the singlet fission sub-cell features a reflective silver back-contact. It is placed at a small off-normal angle from the incoming light, such that light passes through the pentacene layer twice, before and after being reflected at the back-contact, and then reaches the silicon solar cell (see inset Figure 4 (B)). Glass-air and ITO-air interfaces as well as parasitic absorption account for approximately 20 % of light losses in the singlet fission device.

The EQE for this configuration is shown in Figure 4 (B). The silicon solar cell, measured through a glass/ITO/Ag mirror (without the active layers) is shown in blue. The singlet fission device clearly adds to the current, especially where it absorbs most strongly (1.85 eV). The pentacene solar cell alone produces around 60% EQE at this photon energy (green trace). Tantalizingly, the EQE of the parallel tandem cell peaks at 106%, due to the very high IQE of the pentacene cell. Above unity EQE would not be possible without the singlet fission carrier multiplication process, and is something that has not been achieved with a two-bandgap solar cell to date.

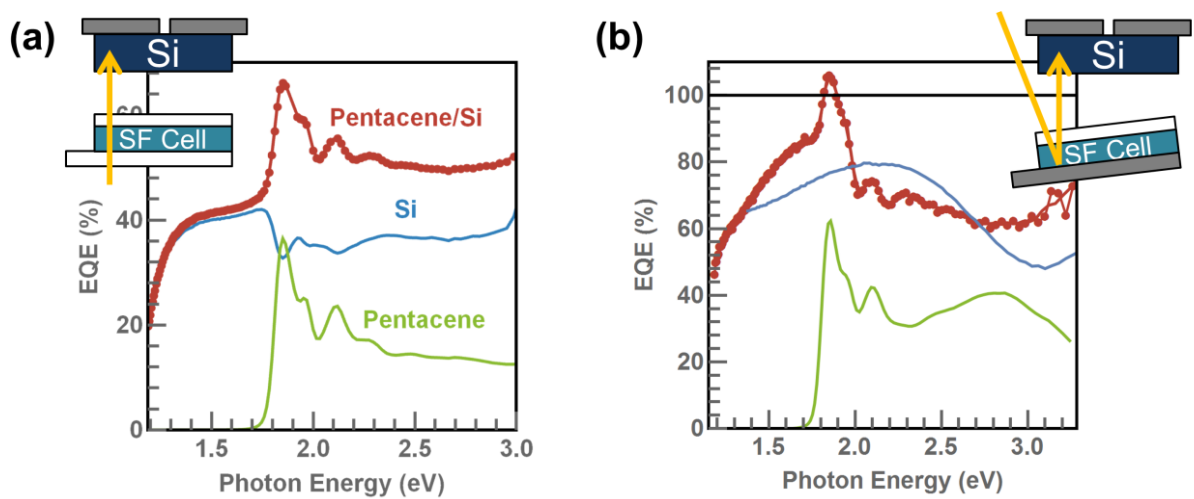

**Figure 4. (A) Pentacene/Silicon tandem cell with the singlet fission sub-cell measured in transmission. With a transparent back contact, transmitted light can pass directly through the singlet fission cell and is absorbed in the silicon cell (see inset). (B) The tandem cell with a reflective silver contact on the pentacene cell, measured in**

**reflection. In this configuration, light passes the singlet fission cell twice before it is reflected into the silicon solar cell.**

For simplicity of construction, we have built the singlet fission solar cell independently from the silicon solar cell and connected both terminals in parallel. For future prospects of this design, the two sub-cells should share a common middle contact to act as a charge-collecting layer, hence reducing manufacturing costs and light absorption in the electrodes. This would be particularly interesting for silicon solar cell configurations that already feature a conductive top contact, such as hetero-junction with intrinsic thin layer (HIT) solar cells. HIT cells currently hold the world record for silicon solar cell efficiency.[9]

In conclusion we have developed a tandem architecture where the two sub-cells are optically connected in series but electrically in parallel. Singlet fission allows doubling the current from high-energy photons and reducing the voltage of the large-bandgap sub-cell to match the voltage of the low-bandgap sub-cell. That way the theoretical efficiency can exceed that of any single-junction solar cell, while being stable against the daily changes of solar spectral irradiance. We have realized a parallel tandem solar cell from the singlet fission sensitizer pentacene together with a monocrystalline silicon solar cell, and demonstrated that the current of the two sub-cells adds. The external quantum efficiency reaches values above 100%, something that would be impossible without the use of carrier multiplication *via* singlet fission

## ACKNOWLEDGEMENTS

The authors thank Erik C Garnett for comments on the manuscript and Moritz Futscher for useful discussions about the efficiency calculations. This work is part of the research program of the Foundation for Fundamental Research on Matter (FOM), which is part of the Netherlands Organization for Scientific Research (NWO) . The authors acknowledge financial support from the Engineering and Physical Sciences Research Council of the UK (EPSRC) and King Abdulaziz City for Science and Technology (KACST). MT acknowledges the Gates Cambridge Trust and the Winton Program for the Physics of Sustainability. The authors declare no competing financial interests. A patent was filed related to this work.

SUPPLEMENTARY INFORMATION FOR

# A Silicon-Singlet Fission Parallel Tandem Solar Cell Exceeding 100 % External Quantum Efficiency

Luis M. Pazos[1,2], Ju Min Lee,[1] Anton Kirch,[2] Maxim Tabachnyk,[2] Richard H. Friend,[2] and Bruno Ehrler[1]

(1)   FOM Institute AMOLF, Center for Nanophotonics, Science Park 104, 1098 XG Amsterdam, The Netherlands
(2)   Cavendish Laboratory, University of Cambridge, JJ Thomson Avenue, Cambridge CB3 0HE, UK

## S1 Methods

**Solar cell fabrication:** The pentacene solar cells were prepared on 12 mm × 12 mm ITO substrates purchased from Psiotec. The ITO substrates were cleaned with acetone and isopropanol by sonification for 10 minutes and oxygen plasma for a further 10 minutes at 250 W. Poly-(3,4-ethylenedioxythiophene):poly(styrenesulfonate) (PEDOT:PSS, Heraeus) was spin coated in air at 4000 rpm before the devices were transferred into a nitrogen filled glovebox. P3HT (Merck) was spin coated from a 4 mg/mL solution in chloroform at 1500 rpm. The devices were then transferred into an evaporation system to deposit the remaining layers at $10^{-6}$ mbar or better. Pentacene (Sigma-Aldrich 698423) and $C_{60}$ (Sigma-Aldrich 572500) were evaporated at 1 Å/s and BCP (Sigma-Aldrich 699152) at 0.5 Å/s. The contact was either a thick (80 nm) silver contact, or a semi-transparent multilayer contact consisting of 1 nm lithium fluoride (Sigma-Aldrich 449903), 1.5 nm aluminum, and 15 nm silver. The solar cells were encapsulated between two glass slides using transparent epoxy.

The silicon solar cell was a commercial Sunpower® solar cell cut into strips of about 2 cm × 12 cm to reduce the background current. The Si solar cells was connected in parallel to the organic solar cell for IV and EQE measurements as described in the main text.

**IV and EQE measurement:** Current-voltage (IV) characteristics were measured in the dark and under a solar simulator (Oriel 92250A) using a Keithley 2636A source-measure unit. The active area was masked for a total area of 4.5 $mm^2$. For external quantum efficiency measurements (EQE) light from a xenon lamp was passed through a monochromator (Oriel Cornerstone 260) onto the solar cell, measured in short circuit. The current from the solar cell was compared to the current of a NIST-traceable calibrated photodiode (Thorlabs SM05-CAL). Both the device and the calibration cell were measured against a reference diode (Thorlabs SM05) to account for changes in light intensity between the measurements. All EQE measurements were taken with chopped monochromatic light under white-light bias. The bias light was generated by a white light-emitting diode reaching an intensity at the solar cell of around 0.3 sun. From sample-to-sample variation and from measurement-to-measurement variation, we estimate the error of the EQE measurement to < 1% relative error in the visible wavelength range (see Figure 1).

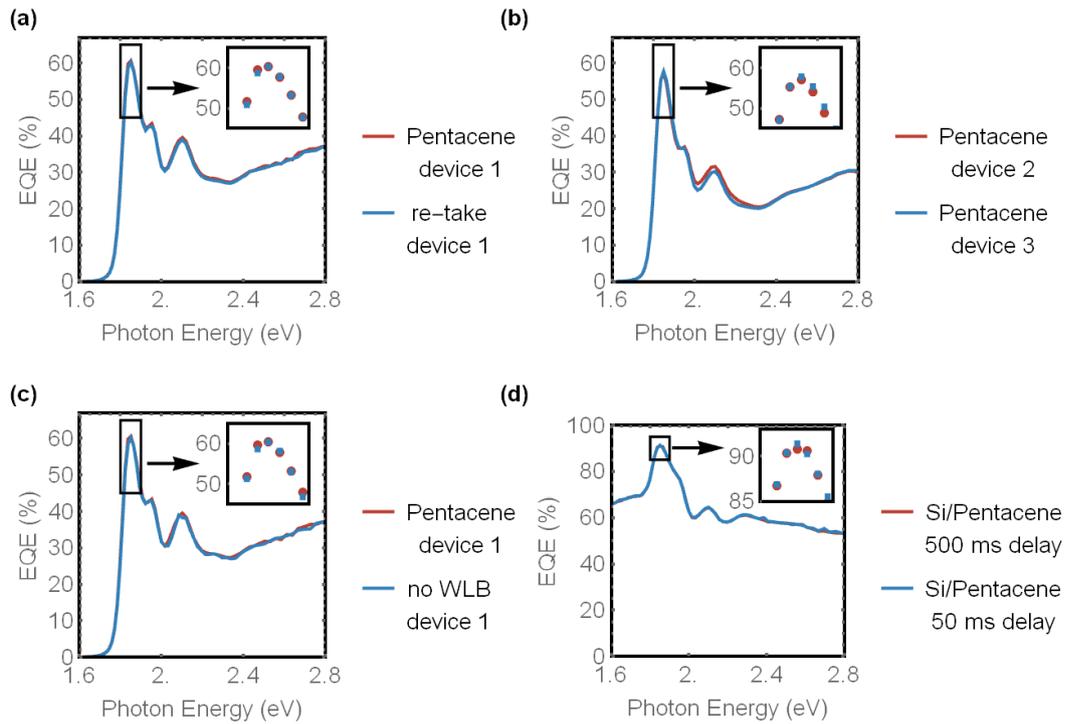

**Figure 1.** (a) Demonstration of the measurement-to-measurement reproducibility, (b) with and without white-light bias (WLB), and (c) Sample-to-sample reproducibility of the pentacene devices. (d) reproducibility of the EQE measurement under varying delay time between taking the data-points.

**Magnetic field dependent measurements:** The devices were placed between the poles of an electromagnet (GMW Model 3470). A cw-diode laser (Thorlabs CPS635S) at 637 nm with an intensity below 10 mW/cm$^2$, chopped at 467 Hz, was used as the pump. The photocurrent was measured with a lock-in amplifier (Stanford SR830). For each magnetic field B, we averaged the B-field response over many cycles. In each cycle, we ramped up the B-field linearly from 0 T over 15 s, then waited 10 s, measured the photocurrent under B-field by averaging over 10 s with 0.5 s between each data point and ramped the B-field linearly down to 0 T over 15 s. Then we waited 10 s and measured the photocurrent with no magnetic field as above.

The photocurrent response was calculated by averaging over 10 cycles. We chose the excitation intensity low to avoid degradation artifacts and made sure that the photocurrent changes by less than 1% during the full measurement.

# S2 Detailed Balance Calculation

The calculation for the maximum theoretical efficiency of our parallel tandem solar cells follows Shockley et al.,[1] with modification for singlet fission similar to Hanna et al.[2]

The generation current of a solar cell is determined by the overlap of the solar spectrum and the quantum yield ($QY$) of charge generation from the solar cell as

$$J_G = \int_{E_G}^{E_{max}} QY(E)\, \Gamma(E) dE$$

where $\Gamma(E)$ is the photon flux, $E_G$ the bandgap of the absorbing semiconductor, and $E_{max}$ the maximum photon energy in the solar spectrum. The maximum quantum yield of a conventional semiconductor is unity above its bandgap ($E_G^{con}$) and zero below, while singlet fission allows for a quantum yield of two from the point where the singlet fission sensitizer absorbs ($E_G^{SF}$). So the generated current in the parallel tandem solar cell from a singlet fission and a conventional semiconductor is

$$J_G = \int_{E_G^{con}}^{E_G^{SF}} \Gamma(E) dE + \int_{E_G^{SF}}^{E_{max}} 2\, \Gamma(E) dE$$

In the ideal situation all recombination current is radiative such that the recombination current can be written as

$$J_R(V) = \frac{2\pi q}{c^2 h^3} \int_{E_G}^{\infty} \frac{QY(E) E^2}{e^{\frac{E - QY(E)\, q V}{kT}} - 1} dE$$

Where $q$ is the elementary charge, $c$ is the speed of light, $h$ is Plank's constant, $k$ is Boltzmann's constant, $T = 300K$ the temperature of the device and $V$ the applied voltage. The recombination current for both cells adds in the parallel tandem architecture, leaving

$$J_R(V) = \frac{2\pi q}{c^2 h^3} \left( \int_{E_G^{con}}^{\infty} \frac{E^2}{e^{\frac{E - q V}{kT}} - 1} dE + \int_{E_G^{SF}}^{\infty} \frac{2\, E^2}{e^{\frac{E - 2 q V}{kT}} - 1} dE \right)$$

Then the conversion efficiency from the solar cell becomes

$$\eta = Max\left(\frac{P_{out}}{P_{sun}}\right) = Max\left(\frac{(J_G - J_R(V))\, V}{P_{sun}}\right)$$

where $P_{out}$ is the power extracted from the solar cell and $P_{sun}$ the incoming power from the sun. Maximizing for voltage allows finding the maximum power converted.

For the series tandem solar cells, generation and recombination current are calculated in the same fashion, but the current matching means that the efficiency is calculated as

$$J_1(V_1) = J_2(V_2)$$

$$J_{G1} - J_{R1}(V_1) = J_{G2} - J_{R2}(V_2)$$

$$V = V_1 + V_2$$

Where $J_1$ and $V_1$ refer to the current and voltage of cell 1 respectively. The two equations are then simultaneously optimized for the product of current and voltage.[2]

Table 1 of the main text shows a few examples for various possible bandgap configurations for single-junction and tandem architectures. Figure 1 (c) and (d) of the main text show the efficiency of a generic series and parallel tandem solar cell with the examples from Table 1 marked.

The spectral data for Figure 2 of the main text was generated from data taken at the Cabauw weather station, which is operated by the Koninklijk Nederlands Meterologisch Instituut (KNMI). The data is deposited on via the World Radiation Monitoring Center Baseline Surface Radiation Network on the Pangaea website. The data used here for the year 2014 can be found in ref. [3–14]. It contains the power received for the direct and diffuse component of the solar radiation. The direct and diffuse part of the AM1.5G spectrum were then scaled to that power and used for the detailed balance calculations as described above.

For Figure 2 (b) of the main text, the efficiency limit was calculated for all spectra of the month of June 2014 and overlaid in the figure. For Figure 2 (c) of the main text, the efficiency limit was first calculated for every spectrum measured at a particular day, and then the efficiency limit was averaged over that day. Each datapoint in Figure 2 (c) represents such an average.

## S3 $V_{OC}$ AND $J_{SC}$ LIMITS COMPARISON

The $J_{SC}$ and $V_{OC}$ for a single-junction solar cell were calculated as in Shockley & Queisser.[1] Figure 2 shows both values normalized to their lowest value in the data range. The $J_{SC}$ decreases by a factor of 9 between 0.8 eV and 2.5 eV, while the $V_{OC}$ changes only by a factor of less than 4.

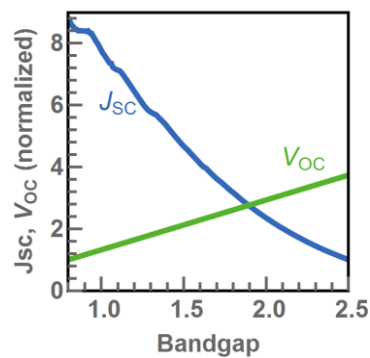

**Figure 2. Limit for the short-circuit current and the open-circuit voltage calculated from detailed balance. Values normalized for the lowest values in the range.**

## S4 EFFICIENCY LIMIT UNDER DIRECT AND DIFFUSE LIGHT

For the efficiency limit calculation under diffuse and direct light the direct and diffuse component of the AM1.5G spectrum as shown in Figure 2(a) of the main text were used to calculate the efficiency limit. The ratio of direct vs. diffuse light was varied such that

$spectrum = ratio \times direct\ spectrum + (1 - ratio) \times diffuse\ spectrum$. Figure 3 shows the efficiency limit as a function of that ratio.

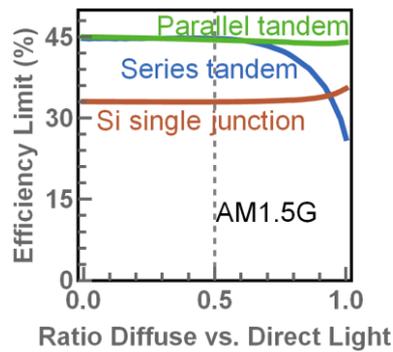

**Figure 3.** Efficiency limit for spectra varied from the AM1.5G spectrum. From only the direct part of the spectrum (0) to only diffuse (1).